		\documentclass{isprs}
		\usepackage{hyperref}
		\usepackage{subfigure}
		\usepackage{setspace}
		\usepackage{geometry}
		 
		\usepackage[utf8]{inputenc}
	
		\usepackage{varioref}

		\usepackage{natbib} 
	
	\usepackage{enumitem}
		\usepackage{graphicx}
		\DeclareGraphicsExtensions{.pdf,.jpg,.png}

		\usepackage[usenames,dvipsnames]{color}

		\usepackage{algorithm2e}
		\usepackage{microtype} 
		
		\usepackage[squaren, Gray, cdot]{SIunits}
		
		\usepackage{listings}
		\usepackage{tabularx}
	
		\usepackage[french,colorinlistoftodos]{todonotes}
		
		\usepackage{setspace}

		\newcommand{\myimage}[3]
					{
					\begin{figure} [h!]
						\begin{center}
							\includegraphics[width=\linewidth,keepaspectratio]{#1}
							\caption{#2}  
							\label{#3}
							\end{center}
					\end{figure} 
					}

		\newcommand{\myimageHL}[4]
		{
			\begin{figure} [ht!]
				\begin{center}
					\includegraphics[width= #4 \linewidth * 2,keepaspectratio]{#1}
					\caption{#2}  
					\label{#3}
				\end{center}
			\end{figure} 
		}

		\newcommand{\myimageFullPageWidth}[3]
								{
								\begin{figure*}[ht]
									\begin{center}
										\includegraphics[width=\textwidth,keepaspectratio ]{#1}
										\caption{#2}  
										\label{#3} 
										\end{center}
								\end{figure*} 
								}

	\usepackage{rotating}

\title{Implicit LOD using points ordering for processing and visualisation in Point Cloud Servers}
\author{Rémi Cura  $^{A}$, Julien Perret $^A$, Nicolas Paparoditis  $^A$}

\address{ $^A$  Université Paris-Est, IGN, SRIG, COGIT \& MATIS, 73 avenue de Paris, 94160 Saint Mandé, France\\
	first\_name.last\_name@ign.fr
	}

\begin{document}
 


\abstract{ 
	Lidar datasets now commonly reach Billions of points and are very dense. 
	Using these point cloud becomes challenging, as the high number of points is untractabel for most applications and for visualisation.
	In this work we propose a new paradigm to easily get a portable geometric Level Of Details (LOD) inside a Point Cloud Server.
	The main idea is to not store the LOD information in an external additional file, but instead to store it implicitly by exploiting the order of the points.
	The point cloud is divided into groups (patches). These patches are ordered so that their order gradually provides more and more details on the patch. 
	We demonstrate the interest of our method with several classical uses of LOD, such as visualisation of massive point cloud, algorithm acceleration,  fast density peak detection and correction. 
}

\maketitle 

\myimageFullPageWidth{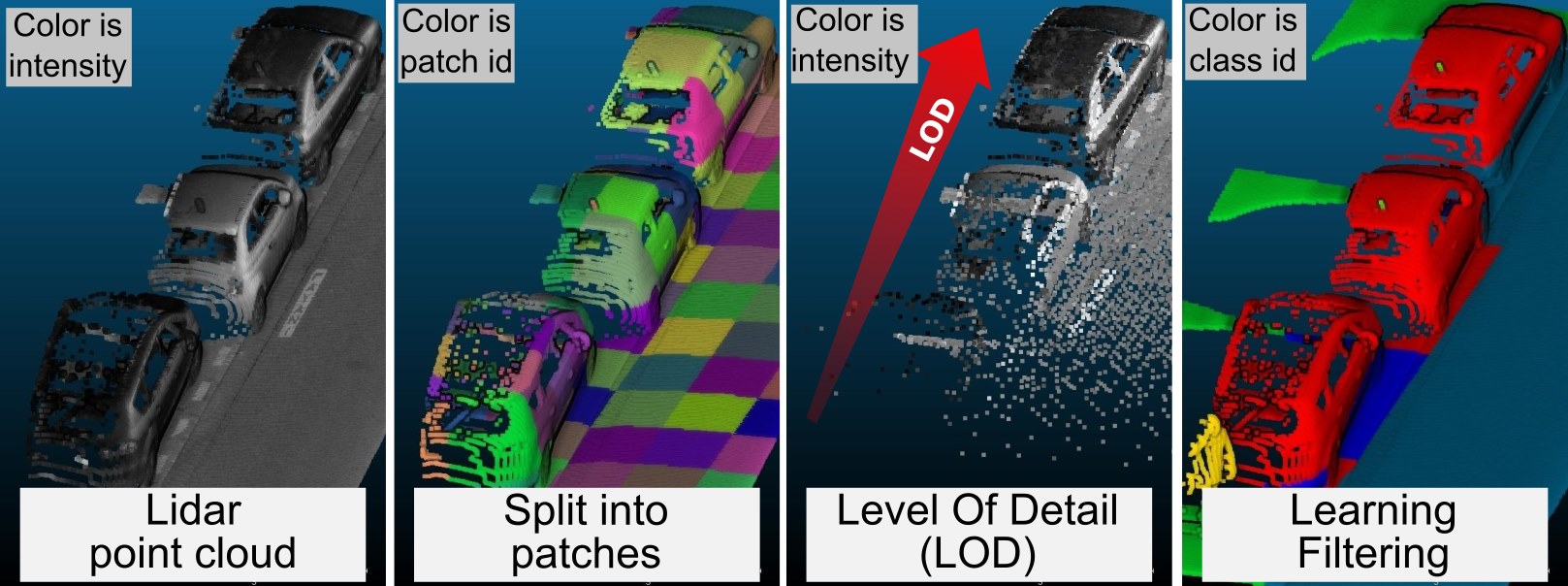}{Graphical Abstract : a Lidar point cloud (1), is split it into patches (2) and stored in a Point Cloud Server, patches are re-ordered to obtain free LOD (3) (a gradient of LOD here), lastly the ordering by-product is a multiscale dimensionality descriptor used as a feature for learning and efficient filtering (4).}{lod.fig:banner_image}



\section{Introduction}

\subsection{Problem}   
	Democratisation of sensing device have resulted into an expansion of acquired point clouds.
	In the same time, acquisition frequency and precision of the Lidar device are also increasing,
	resulting in an explosion of number of points.

	Datasets are now commonly in the multi billion point range, leading to practical issue to store and use them.
	Moreover, point cloud data usage is more common and no more limited to a specialized community. 
	Non specialised users require easy access to data.
	By necessitating easy access and storage and processing for a large amount of data, point clouds are entering the Big Data realm.
	
	Yet all those data are not always needed; having the complete and fully detailed point cloud is impracticable, unnecessary, or even damageable for most applications.
	Therefore, the ability to reduce the number of points is a key point for practical point cloud management and usages.
	
	The number of points must not only be reduced, but often the density corrected.
	Indeed, point clouds from Lidar do not have a constant density.
	The sensing may be structured for the sensing device (for instance a Lidar may sense point using a constant angle), but not necessary for the sensed object (see Fig. \ref{lod.fig:irregular_sampling}). Furthermore,fusing multiple point clouds also produce non regular density.
	\myimageHL{"./illustrations/chap2/problem_in_sampling/regular_vs_irregular_sampling"}{Regular sensing does not imply regular sampling.}{lod.fig:irregular_sampling}{0.5}.
	
	There are basically two approaches to reduce the amount of data considered (See Figure \ref{lod.fig:two_reduction}).
	The first is to use a \textbf{filtering} strategy based on data characteristics (position, time, semantic, etc.) which keeps only a portion the original data.
	The second is a \textbf{generalisation} strategy, where we replace many points with fewer objects that represent appropriately those points. 
	For instance, in order to visualize massive point cloud, it's important to fetch only the appropriate points by selecting the ones which are visible (filtering) and which are the most representative of the scene (generalisation) at the same time.
	 
	Many methods perform filtering, usually by using simple spatial criteria (for instance, points in polygon).
	Generalisation is also popular in its most basic form (generalise points by points).
	\cite{Cura2015} covers extensively filtering with many possibilities (spatial, semantic, attributes, using vector and raster data, using metadata), and also proposes generalisation.
	Nevertheless it uses a generalisation approach only based on more abstract types (bounding box, planes, stats, etc.), which limits its use to methods that are adapted to those types.
	It does not generalise points by points.
	 
	In this work we propose to extend the PCS to explore the generalisation of groups of points by choosing a representative subset of points (See Fig. \ref{lod.fig:two_reduction}).

	We propose to use Level Of Details that reduce successively the number of points while preserving the geometric characteristics of the underlying sensed object.
	Our method is designed to be efficient, robust to point density variation and can be used for many large point clouds processing, including visualisation.
	
	\myimage{"./illustrations/chap2/two_reduction_strategy/two_reduction_strategy"}{Two strategies to limit the amount of points to work on.}{lod.fig:two_reduction}

\subsection{Related Work} 

	Finding a subset of point that represents well all the points is a common problem.
	It has been extensively studied in Geographical Information System (GIS)and other research field.
	It could be seen as compression, clustering, dimensionality reduction, or Level Of Detail (LOD) for visualisation.

	Sophisticated methods have been proposed to generalise 2D points for cartographic applications (\cite{Sester2001}, \cite{Schwartges2013}).
	Yet those methods are limited to 2D points, and could not be easily modified to work in 3D.
	Indeeed, those methods are cartographic by nature, which means that they rely on having all the points on a simple surface : the 3D plan formed by the map.
	Applying directly such methods to point clouds would thus require to have access to surfaces of sensed objects.
	Yet, getting this surface (reconstruction) is a very hard challenge, sometime without solution, and thus we can not rely on it.
	For those limitation and large computing cost, those advanced methods can not be used for large 3D point clouds.

	Other much simpler methods have been designed to work on 3D points. 
	Because the goal is to produce hierarchical levels of points, it seems natural to use a hierarchical structure to compute those levels.
	The main idea is to build a hierarchy of volumes, then each level of the hierarchy corresponds to a LOD. For each volume, a point is created/chosen to generalise all the points within the volume.
	\cite{Rusinkiewicz2000} use a Bounding Sphere Hierarchy for a visualisation application.
	Yet spheres are not well adapted to represent planes, which form a large part of man-made objects and structures.
	On the other hand, Octree (\cite{Meagher1982}) have become the de-facto choice.
	It seems that the most popular use of Octree is as spatial acceleration structure (spatial index).
	Octree have several advantages.
	The first is that their basic nature is closely related to Morton (or GeoHash) order,
	making them efficient to build (\cite{Sabo2014}, \cite{Feng2014}).
	They can also be created out of memory for extremely large point clouds (\cite{Baert2014}). 
	Moreover, their regularity allows efficient representation and compression (\cite{Schnabel2006,Huang2006}), as well as fast geospatial access to data (\cite{Elseberg2013}).

	Octree are also natural candidates to nesting (i.e. create a hierarchy of octrees with various resolution and occupancy, as in \cite{Hornung2013}).
   	Octree construction into file system hierarchy approach is still popular today (\cite{OscarMartinez-Rubi2015}), with point cloud in the 600 Billions points range.
   	It has also been adapted to distributed file system (cloud-computing)  \footnote{\url{https://github.com/connormanning/entwine}}, with processing of 100 Billions points at 2 Billions pts \per hour using a 32 cores 64 GB computer.

	However, the method using Octrees present several disadvantages. 
	Each method uses a custom octree format that is most often stored in an external file.
	This raises problems of concurrency and portability. 

	There a several ways to use an Octree to generalise points.
	We could not find a study of those ways for 3D points. 
	However, \cite{Bereuter2015} recently gave an overview of how quad tree can be used for point generalisation.
	Quad trees are 2D Octrees, yet \cite{Bereuter2015} analyse can be directly translated in 3D. 
 
	The steps are first to compute a tree for the point cloud.
	Then, the point generalisation at a given level is obtained for each cell of the same tree level, by having one point represent all the points of this cell.
	
	There are two methods to choose a point representing the others. The first one is to select on points among all ('select').
	The second method is to create a new point that will represent well the others ('aggregate'). 
	Both these methods can use geometry of points, but also other attributes.
	
	In theory, choosing an optimal point would also depend on application.
	For instance lets consider a point cloud containing a classification, and suppose the application is to visually identify the presence of a very rarely present class C.
	In this case a purely geometrical LOD would probably hide C until the very detailed levels. On the opposite, preferring a point classified in C whenever possible would be optimal for this application.
	
	However, a LOD method has to be agnostic regarding point clouds,
	and point clouds may have many attributes of various type and meaning, as long as many applications.
	Therefore, most methods use only the minimal common factor of possible attributes, that is spatial coordinates. 
	For visualisation applications, aggregating points seems to be the most popular choice \cite{Schutz2015,Hornung2013,Elseberg2013}. with aggregating functions like centroids of the points or centroid of the cell.
	
	All of this methods also use an aggregative function (barycenter of the points, centroid of the cell) to represent the points of a cell.
	Using the barycenter seems intuitive, as it is also the point that minimize the squared distance to other points in the cell, and thus a measure of geometric error.
	
	However, using the 'aggregate' rather than 'select' strategy necessary introduces aggregating errors
	 (as opposed to potential aliasing error), and is less agnostic.
	Indeed, aggregating means fabricating new points, and also necessitate a way to aggregate for each attributes, which might be complex (for instance semantic aggregating; a point of trash can and a point of bollard could be aggregated into a point of street furniture).
	This might not be a problem for visualization application.
	Yet our goal is to provide LOD for other processing methods, which might be influenced by aggregating errors.
	Furthermore, the barycenter is very sensible to density variations.
	
	Therefore, we prefer to use a 'select' strategy. The point to be selected is the closest to the centroid of the octree cell.
	If the point cloud density is sufficient this strategy produces a nearly regularly sampled point cloud, which might be a statistical advantage for processing methods. 
	To establish a parallel with statistics, picking one point per cell is akin to a Latin Hypercube (see \cite{McKay1979}).
	Avoiding the averaging strategy might also increase the quantity of information than can be retrieved (similar to compressed sensing, see \cite{Fornasier2010}).

	We note that most of the LOD systems seems to have been created to first provide a fast access to point (spatial indexing), and then adapted to provide LOD.
	Using the PCS, we can separate the indexing part, and the LOD scheme. From this stems less design constraints, more possibilities, and a method than is not dedicated to only one application (like visualisation).

\subsection{Contribution}

	This work re-uses and combines existing and well established methods with a focus on simplicity and efficiency. As such, all the methods are tested on billions scale point cloud, and are Open Source for sake of reproducibility test and improvements

	\begin{itemize}
			\item   In (Section \ref{lod.method.order}) is to store the LOD implicitly in the ordering of the points rather than externally, avoiding any data duplication.
			Thus, we do not duplicate information, and the more we read points, the more precise of an approximation of the point cloud we get. Reading all the points retrieve the original point cloud.
			
			\item  We introduce (MidOc, Section \ref{lod.method:midoc}), a simple way to order points in order to have an increasingly better geometric approximation of the point cloud when following this order.
		 
	\end{itemize}

\subsection{Plan}
	This work follows a classical plan of Introduction Method Result Discussion Conclusion (IMRAD).
	Section~\ref{lod.sec:method} presents the LOD solution. 
	Section~\ref{lod.sec:result} reports on the experiments validating the methods.
	Finally, the details, the limitations, and potential applications are discussed in Section~\ref{lod.sec:discussion}.



\section{Method}
	\label{lod.sec:method}
	
	In this section, we first present the Point Cloud Server (section \ref{lod.method.PCS})(PCS \cite{Cura2015})
	that this article extends. Then we introduce the LOD solution that we propose 
	, which consists of reordering groups of points from less to more details (\ref{lod.method.order}), and then choose which LOD is needed.
	Although any ordering can be used, we propose a simple geometric one (\ref{lod.method:midoc}) which is fast and robust to density variation. 
	
	\subsection{The Point Cloud Server}
	\label{lod.method.PCS}
		\myimage{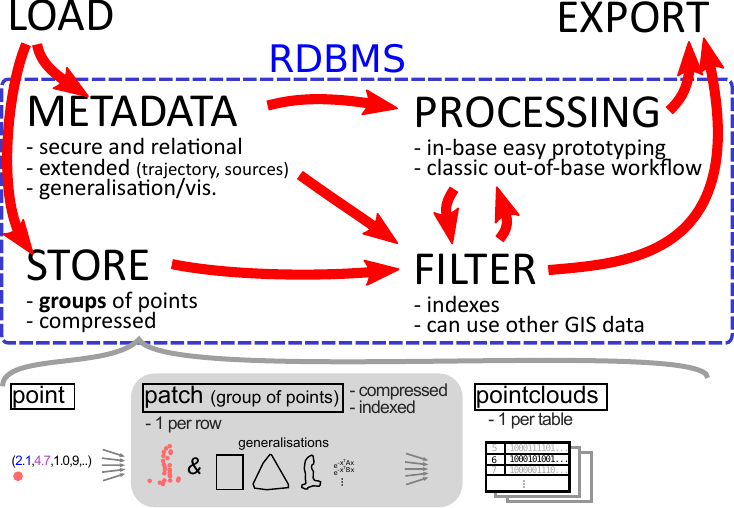}{Overall and storage organisations of the Point Cloud Server.}{lod.fig:PCS}
		
		Our method strongly depends on using the Point Cloud Server described in \cite{Cura2015},
		therefore we introduce its principle and key relevant features (see figure \ref{lod.fig:PCS}).
		
		The PCS is a complete and efficient point cloud management system based on a database server that works on groups of points rather than individual points.
		This system is specifically designed to solve all the needs of point cloud users:
		fast loading, compressed storage, powerful filtering, easy data access and exporting, and integrated processing.
		
		The core of the PCS is to store groups of points (called patches) that are multi-indexed (spatially, on attributes, etc.), and represented with different generalisation depending on the applications.
		Points can be grouped with any rules.
		In this work, the points are regrouped spatially by cubes $1 \metre$ (Paris) or $50 \metre$ (Vosges) wide.
		
		All the methods described in this work are applied on patches.
		We propose is to reorder each patch following the MidOc ordering, allowing LOD and producing a dimensionality descriptor per patch. It can then be used to classify patches.
		
		We stress that our method used on any point cloud will provide LOD,
		but that using it with the PCS is much more interesting,
		and adds key feature such as spatial indexing, fast filtering, etc.

	\subsection{Exploiting the order of points}
		\label{lod.method.order}
			
			\myimageHL{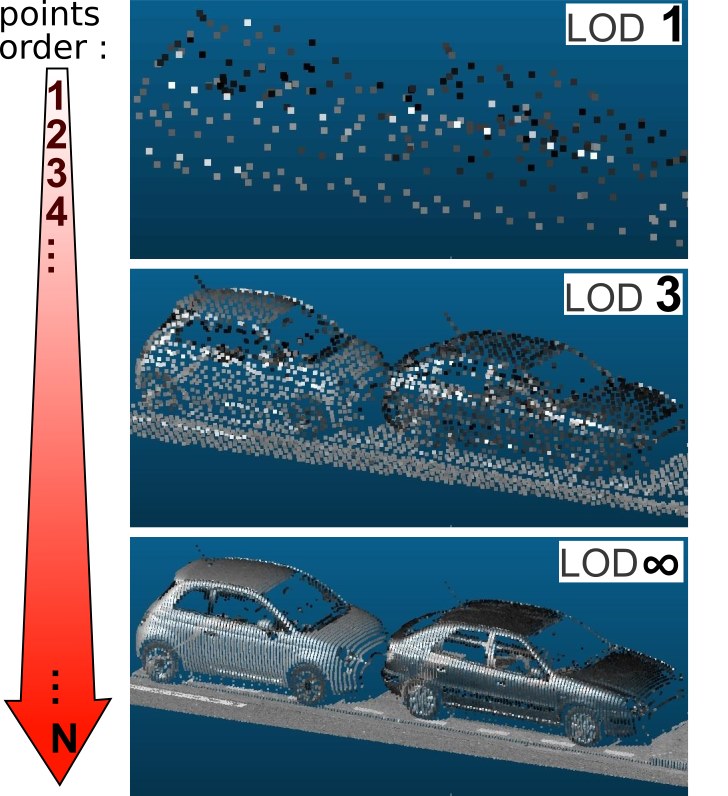}{3 Geometrical Level Of Detail (LOD) for the same point cloud. Reading points from 1 to N gradually increases the details, because of the specific order of points (MidOc).}{lod.fig:lod-principle}{0.5}

			We propose to exploit the ordering of points to indirectly store LOD information.
			Indeed, whatever the format, be it file or database based, points ends up as a list, which is ordered.
			 
			The idea is then to exploit the order of this list, so that when reading the points from beginning to end, we get gradually a more accurate geometrical approximation of the point cloud (see figure \ref{lod.fig:lod-principle}). 
			 
			For instance, given list $L[P_1,..,P_N]$ of ordered points.
			Reading $P_1$ to $P_5$ gives a rough approximation of the point cloud, and reading another $16$ points ($P_1$ to $P_{21}$) is going to give a slightly better approximation. Reading points $1$ to $N$ is going to get the exact point cloud, so there is no data loss, nor data duplication.
		 
			Using the point ordering as LOD results in three main advantages.
			\paragraph{Implicit}
			Except a pre-processing step to write the point cloud following a given ordering, each time the user wants to get a Level Of Detail version of the point cloud, there is no computing at all (only data reading).
			This may not make a big difference for non-frequent reading, but in a context where the same point cloud is going to get used several times at several levels and by several users simultaneously (for instance Point Cloud as a Service), no processing time makes a big difference.
		
			\paragraph{No Duplication}
			Another big advantage is that exploiting point ordering does not necessitate additional storage.
			This is an advantage on low level. It saves disk space (no data duplication, no index file). Because the LOD information is embedded right within the point cloud, it is perfectly concurrent-proof, i.e. the point cloud and the LOD can not become out of sync.
			(Even in heavy concurrent Read/Write, a user would always get a coherent LOD).
			Lastly because the LOD only relies on ordering the original points, and does not introduces any other points or data, it avoids all precision-related issues that may come from aggregating.
			
			\paragraph{Portable}
			The last advantage comes from the simplicity of using the ordering. 
			Because it is already something that all point cloud tools can deal with (a list of points!), it is portable. Most softwares do not change the points order inside a cloud (See Section \ref{lod.result.os_softwares}).
			Even if a tool were to change the order, it would be easy to add the ordering number as an attribute (though slightly increasing the storage requirement).
			This simplicity also implies that adapting tools to exploit this ordering is very easy.

	\subsection{MidOc : an ordering for gradual geometrical approximation}
		\label{lod.method:midoc}
		\subsubsection{Requirements and hypothesis}
		\label{lod.method.midoc.hypothesis}
		The method exploits the order of points to store LOD information, so that the more points are read, the more detailed the result becomes.
		Obviously an ordering method that class the points from less details ($LOD_0$) to full details($LOD_\infty$) is needed.
		This ordering is in fact a geometric measure of point relevance, that is how well a point represents the point cloud (in a neighbourhood depending of the LOD).
		
		This ordering will be used by on different point clouds and for many applications, and so can not be tailored to one.
		As such, we can only consider the geometry (the minimal constituent of a point).
		Because of the possible varying-density point clouds, the ordering method also have to recreate a regular-ish sampling.
		
		Although many ordering could be used (for example, a simple uniform-random ordering),
		a suitable one would have low-discrepancy (that is be well homogeneous in space, see \cite{Rainville2012}), not be sensitive to density variations, be regular, be fast to compute and be deterministic (which simplify the multiuser use of the point cloud).
		
		We make two hypothesis that are mostly verified on Lidar point cloud.
		The first hypothesis ('disposable density') is that the density does not gives information about the nature of the object being sensed. 
		That is, depending on the sensing situation, some parts of the cloud are more or less dense, but this has nothing to do with the nature of the object sensed, thus can be discarded.
		The second hypothesis (low noise) is that the geometrical noise is low.
		We need this hypothesis because 'disposable density' forbids to use density to lessen the influence of outliers.
		
		A common method in LOD is to recursively divide a point cloud into groups and use the barycentre of the group as the point representing this group. The ground of this method is that the barycentre minimise the sum of squared distance to the points.
		
		However such method is extremely sensible to density variation, and artificially creates new points. 
		
		\subsubsection{Introducing the MidOc ordering}
		
		\myimage{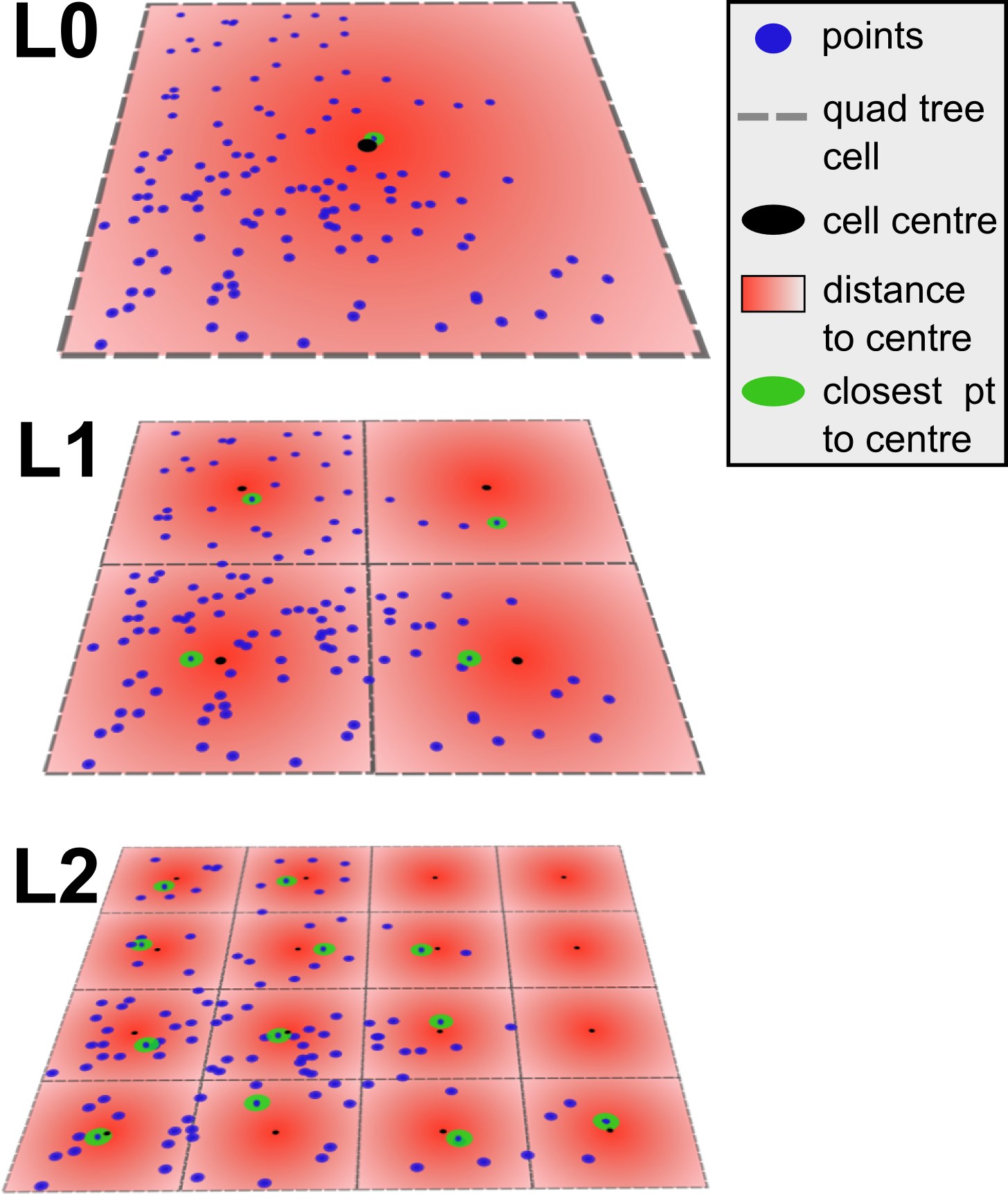}{MidOc explained in 2D. Given a point cloud (Blue) and quad tree cells (dashed grey), the chosen point (green ellipse) is the one closest to the centre (black point) of the cell.}{lod.fig:midoc-principle}{0.75}
		
		We propose the re-use of well known and well proven existing methods that is the octree subsampling (for instance, the octree subsampling is used in \cite{Girardeau-Montaut2014}).
		An octree is built over a point cloud, then for each cell of the octree the LOD point is the barycentre of the points in the cell.  With this, browsing the octree breadth-first provides the points of the different levels.
		
		We adapt this to cope with density variation, and to avoid creating new point because of aggregation.   
		We name this ordering MidOc (Middle of Octree subsampling) for clarity, nonetheless we are probably not the first to use it.
		
		The principle is very simple, and necessitate an octree over the point cloud (octree can be implicit though).
		We illustrate it on Figure \ref{lod.fig:midoc-principle} (in 2D for graphical comfort).
		We walk the octree breadth-first.
		For each non-empty cell, the point closest to the cell centre is chosen and assigned the cell level,
		and removed from the available point to pick.
		The process can be stopped before having chosen all possible points,
		in which case the remaining points are added to the list, with the level $L_\infty$.
		
		The result is a set of points with level $(P,L_i)$.
		Inside one level $L_i$, points can be ordered following various strategies (see Section \ref{lod.method.intralevel}).
		
		\subsubsection{Implementation}
		\label{lod.method.midoc.implementation}
		MidOc ordering is similar to octree building. Because Octree building has been widely researched, we test only two basic solutions among many possibilities.
		
		The first kind of implementation uses SQL queries. For each level, we compute the centres of the occupied cells using bit shifts and the closest point to these. Picked points are removed, and the process is repeated on the next level.
		It relies on the fact that knowing each point octree cell occupancy does not require to compute the octree (see Figure \ref{lod.fig:binary_coordinates_example}).
		
		The second implementation uses python with a recursive strategy. it only necessitates a function that given a cell and a list of points chose the point closest to the centre of the cell, then split the cell and the list of points for the next level, and recursively calls itself on this subcells with the sublists.
		
		A more efficient and simpler implementation is possible by first ordering the points following the Morton (Hypothesis : or Hilbert) curve, as  in \cite{Feng2014} (Section 2.5.1, page 37), in the spirit of linear octree.
		
		\subsubsection{Intra-level ordering}
		\label{lod.method.intralevel}
		
		\myimage{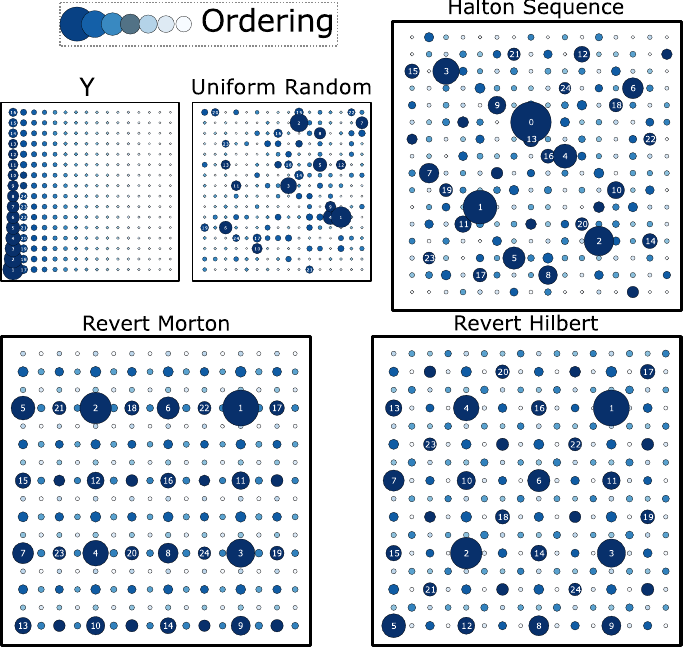}{Several possible intra-level orders with various coverage from bad to good. Revert Morton and Revert Hilbert have offset for illustration.}{lod.fig:intralevel_ordering}
		
		Inside one LOD points can be ordered with various methods.
		The intra-level ordering will have an impact if the LOD is used in a continuous way,
		and moreover may influence methods that relies on low-discrepancy.
		More precisely, if only a part of the points in a level are going to be used,
		it may be essential that they cover well the spatial layout of the totality of points.
		Several methods give this kind of coverage (see \cite{Rainville2012})
		
		Lets take the example where the goal is to find the plan that best fits a set of points
		and the method to do so in online (for instance it could be based on online robust PCA like in (\cite{Feng2013})).
		The plan fitting method reads one by one the points of a designated level $L_i$, and successively computes a better plan estimation.
		
		The Figure \ref{lod.fig:intralevel_ordering} presents some possible ordering. 
		If the plan detection method was applied on the Y ordering, it would necessitate a great number of points to compute a stable plan. For instance the first 16 points (1 column) would not permit to compute a plan.
		Similarly, if the point were ordered randomly, estimating a plan would still require lots of points, because uniform randomly chosen points are not well spread out (on the figure, the first 25 points are over represented in the upper left part).
		
		On the opposite, using a low discrepancy ordering like the Halton sequence makes the points well spread, while being quasi-random.
		Inverted space filling curves like the Morton or Hilbert curves also cover well space, at the price of being much more regulars.
		
		The Halton sequence ordering is obtained by generating a Halton sequence (nD points) and successively pick points closest to the Halton generated points.
		The revert Morton ordering and revert Hilbert ordering are the distance along Morton or Hilbert curve expressed in bit and read backward (with a possible offset).

	    \subsubsection{Points streaming from the PCS for interactive web-based visualisation}
		The open source project \\
		LOPOCS\footnote{\url{https://li3ds.github.io/lopocs/}} developped by Paul Blottiere (Oslandia) implements the LOD concepts and propose a WebGL-based prototype for visualisation.
		
		The number of points per patch sent to the browser is limited using LOD. Patch are ordered with MidOc, so the visual artefact is greatly reduced, and the data loads more quickly, as expected. 
		
		We also use MidOc at the table level to reduce the number of patches used for visualisation.
		Indeed, simply using MidOc at the patch level results in minimum to one point per patch. 
		Yet many patches may be in the view frustrum, which would severely affect performances.
		
		Using MidOc at the table level, only the most relevant of patches which are generalised with only one point are selected.
		
		The global architecture of LOPOCS uses a web server between the Point CLoud Server and the client browser, enabling asynchronous loading of points.
		
		LOPOCS determines desired LOD of each patch based on a classical view-frustrum method which amount to how much screen space the patch bounding box will occupy.
				 
	\subsection{Excessive Density detection and correction}
		\label{lod.method.density}
		Lidar point cloud do not have a constant density, even if the acquisition is performed at a constant sensing rate, because the sensed object geometry (See Fig. \ref{lod.fig:irregular_sampling}).
		
		Important variation of density can be a serious issue for some processing methods. 
		For instance if millions of points are concentrated in a small volume,
		a processing method operating on fixed size volume may exceed the maximum memory of the system.
		Large density variation are also bad for performances in parallel environment.
		Indeed, efficient parallel computing may require that all the workers have about the same amount of work.
		One worker stumbling upon a very dense part of the point cloud would have much more points to process than the other workers.
		The figure \ref{lod.fig:density-correction} shows a place in the Paris dataset where the density is 5 times over the average value of this data set.
		In this context of terrestrial Lidar, this density peak is simply due to the fact that the acquisition vehicle stopped at this place
		, while continuing to sense data.
		
		The PCS coupled with LOD patches allows to quickly find abnormally high density.
		The PCS filters in few milliseconds the patch containing lots of points. This suffice for most applications.
		For a finer density estimation, we compute the approximate volume of the patch.
		For a level $L$, the $ppl[L]$ number of points multiplied by the theoretical cell size for this level gives an approximate volume (or surface) of the patch.
		The total number of points divided by this volume (surface) gives a finer volumetric (surface) density estimation.
		
		Then, correcting density consists of taking into account only the first $K$ points, where $K$ is computed to attain the approximate patch volume (surface).

	

 \section{ Result }
	 \label{lod.sec:result}
 	\subsection{Introduction to results}
 		We design and execute several experiments in order to validate all points  introduced in Section \ref{lod.sec:method}.
 		First we prove that is it effectively possible to leverage points order, even using canonical open sources software out of the box.
 		Second we perform MidOc ordering on very large point cloud and analyse the efficiency, quality and applications of the results.
 		 
 		The base DBMS is \cite{PostgreSQL2014}. The spatial layer \cite{PostGIS2014} is added to benefits from generic geometric types and multidimensional indexes. The specific point cloud storage and function come from \cite{pgPointCloud2014}. 
 		The MidOc is either plpgsql or made in python with \cite{SciPy2014}.  
 		Timings are only orders of magnitude due to the influence of database caching.
 	
		\myimageHL{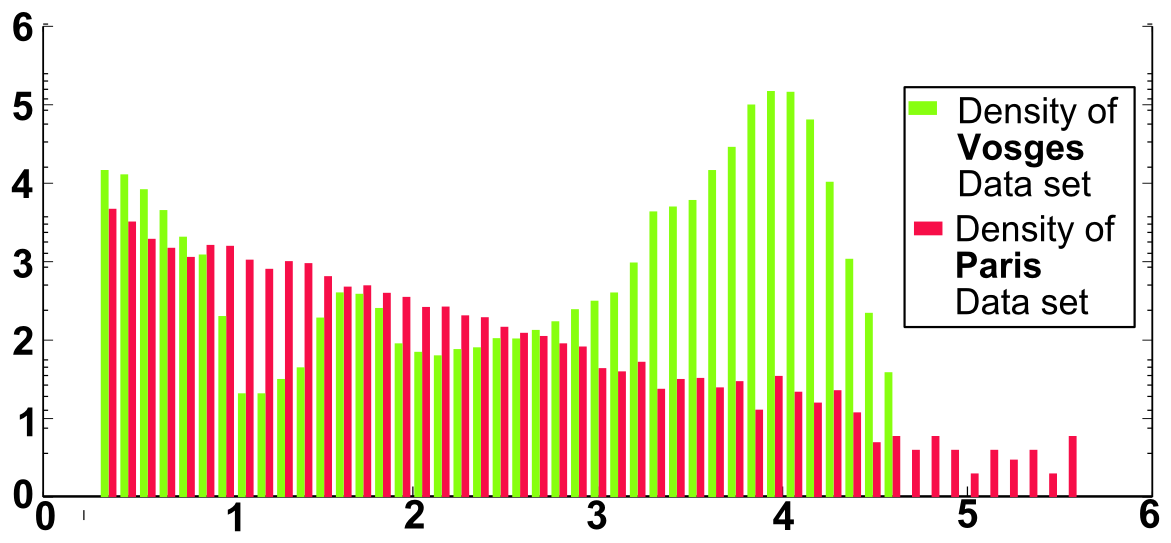}{ Histogram of number of points per patch, with a logarithmic scale for X and Y axis}{lod.fig:hist-density-dataset}{0.5}
	 
 		We use two data sets. There were chosen as different as possible to further evaluate how proposed methods can generalise on different data (See Figure fig:hist-density-dataset for histogram of patch density ). 
 		The first data set is \cite{IQmulus2014} (Paris data set), an open source urban data set with varying density, singularities, and very challenging point cloud geometry. 
 		Every point is labeled with a hierarchy of 100 classes.
 		The training set is only 12 millions points.
 		Only 22 classes are represented. We group points in $1 \cubic \metre$ cubes.
 		The histogram of density seems to follow an exponential law (See figure \ref{lod.fig:hist-density-dataset}), the effect being that many patches with few points exist. 
 		
 		We also use the Vosges data set, which is a very wide spread, aerial Lidar, 5.5 Billions point cloud. 
 		Density is much more constant at 10k pts/patch .
 		A vector ground truth about surface occupation nature (type of forest) is produced by the French Forest Agency. Again the classes are hierarchical, with 28 classes.
 		We group points in 50 $\times 50 \meter$ squares.

	\subsection{Using the Point Cloud Server for experiments}
		All the experiments are performed using a Point Cloud Server (cf \cite{Cura2014}).
		The key idea are that point clouds are stored inside a DBMS (postgres), as patch. Patch are compressed groups of points along with some basic statistics about points in the group.
		We hypothesize that in typical point cloud processing workflow, a point is never needed alone, but almost always with its surrounding points.
	
		Each patch of points is then indexed in an R tree for most interesting attributes (obviously X,Y,Z but also time of acquisition, meta data, number of points, distance to source, etc.)
			
		Having such a meta-type with powerful indexes allows use to find points based on various criteria extremely fast. (order of magnitude : ms). 
		As an example, for a 2 Billion points dataset, we can find all patches in few milliseconds having : 
		 - between -1 and 3 meters high in reference to vehicle wheels
		 - in a given 2D area defined by any polygon 
		 - acquired between 8h and 8h10 - etc.
		 
		The PCS offers an easy mean to perform data-partition based parallelism. We extensively use it in our experiments.

	\subsection{Exploiting the order of points}
		\label{lod.result.os_softwares}
		We proposed to implicitly store LOD in the order of the points (Section \ref{lod.method.order}).
		In this first experiment we check that point cloud ordering is correctly preserved by common open source point cloud processing software.
		For this, we use a real point cloud, which we order by MidOc ordering. 
		We export it as a text file as the reference file.
		For each software, we read the reference file and convert it into another format, then check that the conversion did not change the order of points. 
		The tree common open source software tested are CloudCompare\footnote{\url{www.danielgm.net/cc}}
		, LasTools\footnote{\url{www.cs.unc.edu/~isenburg/lastools}} and Meshlab\footnote{\url{http://meshlab.sourceforge.net/}}.
		All pass the test.
			
	\subsection{MidOc: an ordering for gradual geometrical approximation}
		\subsubsection{MidOc at the patch level}
		\myimageFullPageWidth{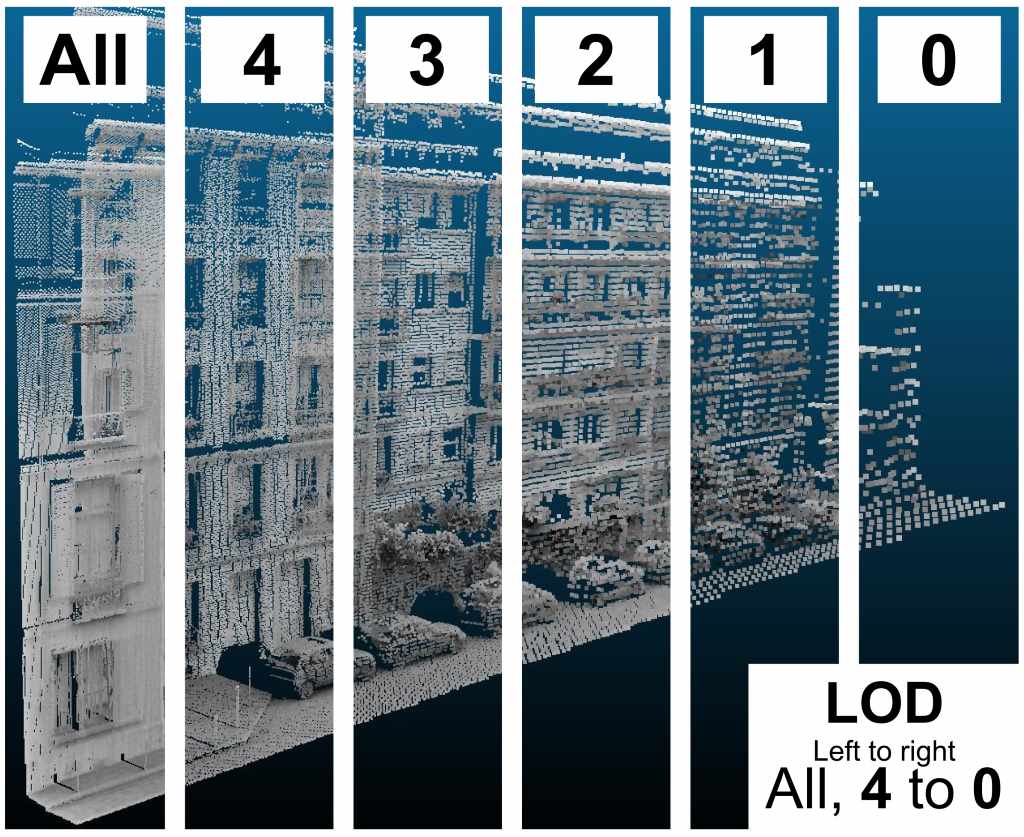}{Schematic illustration of different LOD. Left to right, all points, then LOD 4 to 0. Visualized in cloud compare with ambient occlusion. Point size varies for better visual result.}{lod.fig:visual_LOD_left_right}

		We first test the visual fitness of MidOc ordering.
		Then we compute MidOc for our two datasets and evaluate the trade-off between point cloud size and point cloud LOD. 
		As a proof of concept we stream a 3D point cloud with LOD to a browser.
		
		The figure \ref{lod.fig:visual_LOD_left_right} illustrates LOD on a typical street of Paris dataset

		We compute the size and canonical transfer time associated for a representative street point cloud.
		For this order of magnitude, the size is estimated at 5*4 Byte (5 floats) per point, and the (internet) transfer rate at 1 \mega $byte$\per \second.
		
		\begin{table}[ht]
			\centering
			\caption{Number of points per LOD for the point cloud in the Figure \ref{lod.fig:visual_LOD_left_right}
				, plus estimated transfer time at 1 \mega $byte$\per \second.}
			\label{lod.tab:lod-size-time}
			\scriptsize 
			\begin{tabular}{cccccc}
				\textbf{Level} & \shortstack{\textbf{Typical} \\ \textbf{spacing (\centi \meter)}} & \shortstack{ \textbf{Points} \\ \textbf{number (k)}} & \shortstack{\textbf{Percent of} \\ \textbf{total size}} & \shortstack{\textbf{Estimated} \\ \textbf{time (\second)}}   \\
				\hline All & 0.2 to 5  & 1600 & 100 & 60 \\ 
				\hline 0 & 100 & 3 & 0.2 & 0.1 \\ 
				\hline 1 & 50 & 11.6 & 0.7 & 0.4 \\ 
				\hline 2 & 25 & 41 & 2.6 & 1.5 \\ 
				\hline 3 & 12 & 134 & 8.4 & 5 \\ 
				\hline 4 & 6 & 372 & 23 & 14 \\    
			\end{tabular} 
		\end{table}
			 
		
		We use 3 implementations of MidOc, two being pure plpgsql (postgreSQL script langage), and one Python (See Section \ref{lod.method.midoc.implementation}).
		We successively order all the Paris and Vosges data sets with MidOc, using 20 parallel workers, with a plpgsql implementation.
		The ordering is successful on all patches, even in very challenging areas where there are big singularities in density, and many outliers.
		The total speed is about 100 millions points/hour using in-base processing.
		We prototyped an out-of-base processing where the extraction of points from patch is done on the client, and reached a 180 \mega pts \per \hour.
		The same method, without any ordering (only converting patch to point then point to patch) reach a 2.3 B pts\per\hour. 
		We consider it to be at least 10 times too slow for practical use.
		We briefly analyse performances, and conclude that only 10 workers are efficient.
		
		\subsubsection{MidOC at the table level}
		When using MidOc at the patch level, a patch will have at least one point when using the coarsest LOD.
		Yet a table may contains millions of patches, which means that using only patch-level LOD, the coarsest LOD could still contain millions of points.
		
		To solve this problem, we introduce MidOc at the table level.
		In the PCS, each patch amount to one row in a point cloud table, and is represented by a point at teh coarsest LOD.
		We order those point using MidOc, and write the ranking in an additional column.
		
		Then the maximum number of patches can be limited simply by adding to the SQL query "ORDER BY midoc LIMIT X", where X is the max number of patches.
		
		This simple mechanism allow an overall control on disk reading from the server, as this is related to number of patch read.

		\subsubsection{Points streaming from the PCS for interactive web-based visualisation}
 
		 Several demonstrations are available\footnote{\url{https://li3ds.github.io/lopocs/}}, using point clouds from 3 to 300 millions points.
		 The asynchronous loading means that the user browser is never frozen while waiting for points.
		 Instead, the user is free to explore, and points are constantly added to the visualisation.
	
		We stress that the streaming approach also heavily relies on patch filtering, as only patch inside the view frustrum are considered, which is a fast spatial query in the PCS.

	\subsection{Excessive Density detection and correction} 
		\myimage{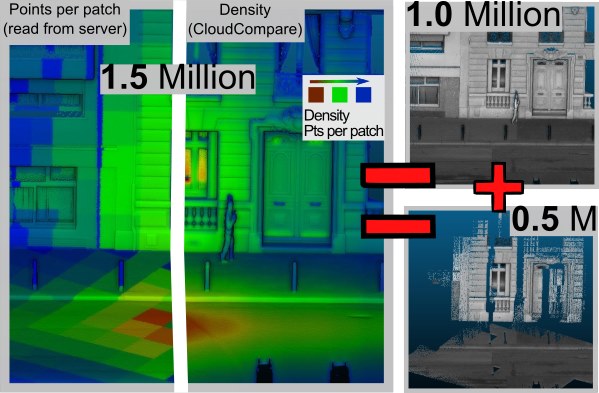}{Abnormal density detection and correction. Top points per patch (left) or density (right), green-yellow-red. Bottom reflectance in grey. }{lod.fig:density-correction}
		
		We detect the abnormal density (explained in Section \ref{lod.method.density}) in the Paris data set in $\sim 100 \milli \second$ 
		(See Figure \ref{lod.fig:density-correction}). 
		In comparison, computing the density per point with neighbourhood is extremely slow (only for this 1.5 Million extract, 1 minute with CloudCompare,4x2.5GHz, 10cm rad) (top right illustration), and after the density is computed for each points, all the point cloud still need to be filtered to find abnormal density spot.
		
		If the patch are ordered following MidOc, unneeded points are removed by simply putting a threshold on points per patch (bottom left, 1 to 5k points \per \cubic \meter , bottom right , 5k to 24 k pts \per \cubic \meter). It considerably reduces the number of points (-33\%). 
		
		This strategy can be automated by stating than no patch should return points over Level $L_i$. Then when getting points from the PCS, so that only points in those levels are sent.



 \section{ Discussion }
	 \label{lod.sec:discussion} 
	 
	 \subsection{Point cloud server}
	\label{lod.par:pointcloudserver-limitation}
	We refer the reader to \cite{Cura2015} for an exhaustive analyse of the Point Cloud Server.
	Briefly, the PCS has demonstrated all the required capacities to manage point clouds and scale well.
	To the best of our knowledge the fastest and easiest way to filter very big point cloud using complex spatial and temporal criteria, as well as natively integrate point cloud with other GIS data (raster, vector).
	The main limitation is based on the hypothesis than points can be regrouped into meaningful (regarding further point utilisation) patches. If this hypothesis is false, the PCS lose most of its interest.

	\subsection{Exploiting the order of points}
	From a practical point of view, implicitly storing the LOD using the point ordering
	seems to be estremely portable. Most softwares would not change the order of points.
	For those who might change the order of points, 
	it is still very easy to add the order as an attribute, thus making it fully portable.
	However, this approach has two limitations.
	The first limitation is that the order of point might already contains precious information. 
	For instance with a static Lidar device,
	the acquisition order allows to reconstruct neighbourhood information.
	The second limitation is that an LOD ordering might conflict with compression.
	Indeed ordering the points to form LOD will create a list of points were successive points are very different. Yet compressing works by exploiting similarities.
	A software like LasTool using delta compressing might suffer heavily from this. 
			 
	 \subsection{MidOc : an ordering for gradual geometrical approximation}
	 	We stress that the LOD are in fact almost continuous (as in the third illustrations of Fig.  \ref{lod.fig:banner_image}). 
		 	
		MidOc is a way to order points based on their importance. In MidOc,
		the importance is defined geometrically.
		Yet specific applications may benefit from using other measure of importance, 
		possibly using other attributes than geometrical one,
		and possibly using more perceptual-oriented measures.
		
		MidOc relies on two hypothesis which might be false in some case. 
		Indeed, variation of density may be a wanted feature
		(e.g. stereovision, with more image on more important parts of the object being sense).
		Again the low geometrical noise hypothesis might be true for Lidar, but not for Stereo vision 
		or medical imaging point cloud. However in those case de-noising methods may be applied before computing MidOc.

	\subsubsection{Applications}
		MidOc ordering might be of use in 3 types of applications. First it can be used for graphical LOD, as a service for point cloud visualisation.
		Second the ordering allows to correct density to be much more constant.
		Complex processing methods may benefits from an almost constant density, or for the absence of strong density variation.
		Third the ordering can be used for point cloud generalisation,
		as a service for processing methods that may only be able to deal with a fraction of the points. 
		
		The illustration \ref{lod.fig:visual_LOD_left_right} gives visual example of LOD result and how it could be used to vary density depending on the distance to camera.  
		It is visually clear that the rate of increase of points from LOD 0 to 4 for floor lamp (1D) window (2D) and tree (3D) is very different.
		Small details are also preserved like the poles or the antenna of the car. preserving those detail with random or distance based subsampling would be difficult.

	\subsubsection{Implementation}
		\label{lod.subsubsec:bit_coordinates}
		Octree construction may be avoided by simply reading coordinates bitwise in a correctly centred/scaled point cloud.
		We centre a point cloud so that the lowest point of all dimension is $(0,0,0)$, and scale it so that the biggest dimension is in $[0,1[$.
		The point cloud is then quantized into $[0..2**L-1]$ for each coordinate.
		The coordinates are now integers, and for each point, reading its coordinates bitwise left to right gives the position of the point in the octree for level of the bit read.
		This means performing this centering/scaling/quantization directly gives the octree. Moreover, further operations can be performed using bit arithmetic, which is extremely fast.
		\myimageHL{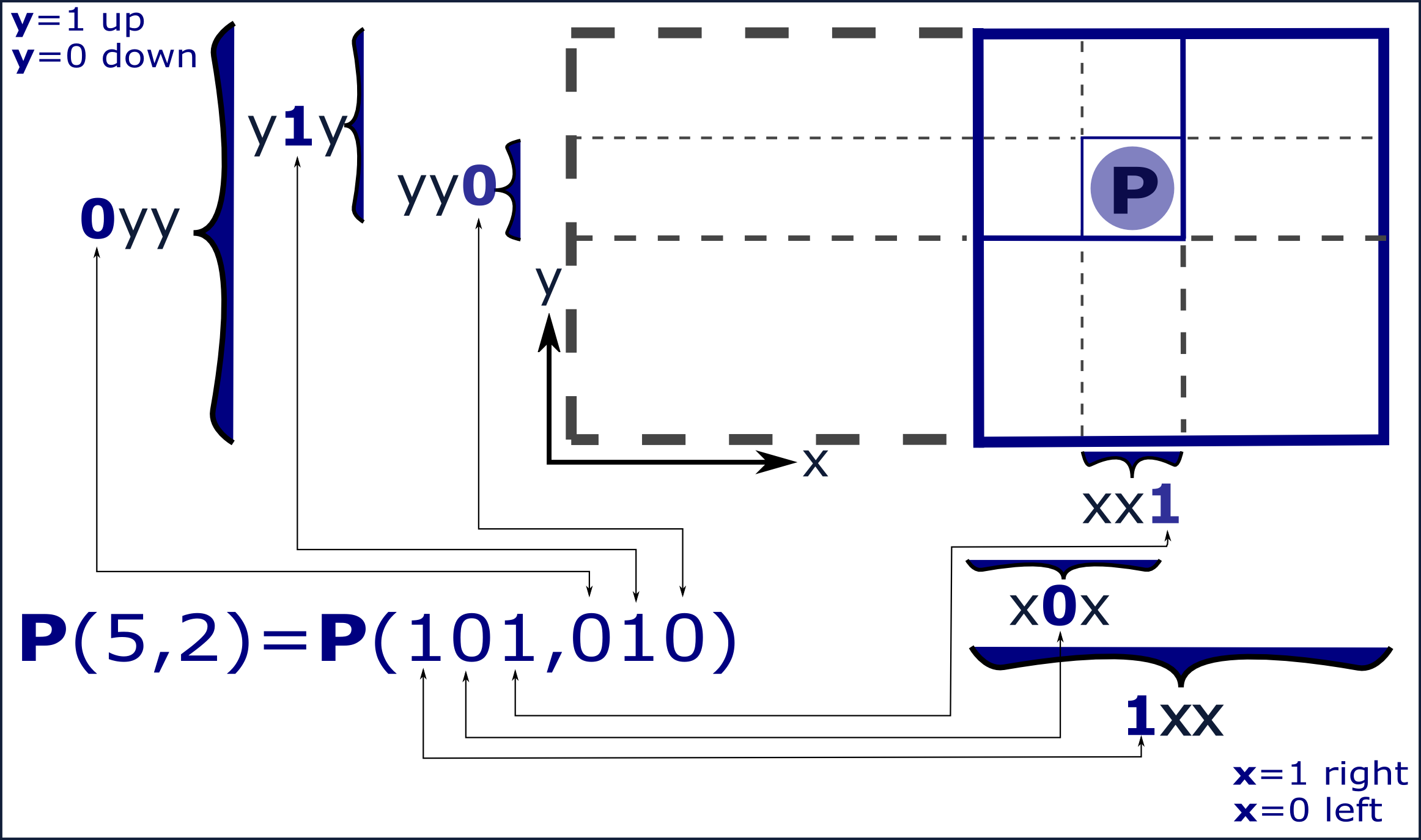}{Principle of binary coordinates for a centered, scaled and quantized point cloud.}{lod.fig:binary_coordinates_example}{0.5}
		
		On this illustration the point $P$ has coordinates $(5,2)$ in a $[0,2^3-1]^2$ system. Reading the coordinates as binary gives $(b'101',b'010')$.
		Thus we know that on the first level of a quad tree, $P$ will be in the right (x=$b'1xx'$) bottom (y=$b'0yy'$) cell.
		For the next level, we divide the previous cell in 2, and read the next binary coordinate. $P$ will be in the left (x=$b'x0x'$) up (y=$b'y1y'$) cell. There is no computing required, only bit mask and data reading.
			
		Regarding implementation, the three we propose are much too slow, by an order of magnitude to be easily used in real situation. We stress however that the slowness comes from inefficient data manipulation, rather than from the complexity of the ordering. 
		It may also be possible to use the revert Hilbert ordering to directly compute MidOc.
		Furthermore, octree construction has been commonly done on GPU for more than a decade.
	 
	 \subsubsection{Size versus LOD trade-off}
		 \label{lod.point-cloud-server-troughput}
		 The table \ref{lod.tab:lod-size-time} shows that using the level 3 may speed the transfer time by a 10 factor.
		 The point cloud server throughput is about 2-3 \mega $byte$ \per \second  (monoprocess),  sufficient for an internet throughput, but not fast enough for a LAN 10 \mega $byte$ \per \second.
		 This relatively slow throughput is due to current point cloud server limitation (cf \ref{lod.par:pointcloudserver-limitation}).
	 
	 \subsubsection{Large scale computing} 
		 The relatively slow computing (180 Millions points \per \hour ) is a very strong limitation.
		 This could be avoided. A C implementation which can access raw patch would also be faster for ordering points.
		 
	 \subsubsection{Points streaming from the PCS for interactive web-based visualisation}
		 Streaming low level of detail patches greatly accelerate visualisation,
		 which is very useful when the full point cloud is not needed.
		 To further accelerate transmission, patch LOD is determined according to the distance to camera (frustrum culling). (See Figure \ref{lod.fig:lod-dist-to-camera} for a naive visual explanation.)
		 \myimageFullPageWidth{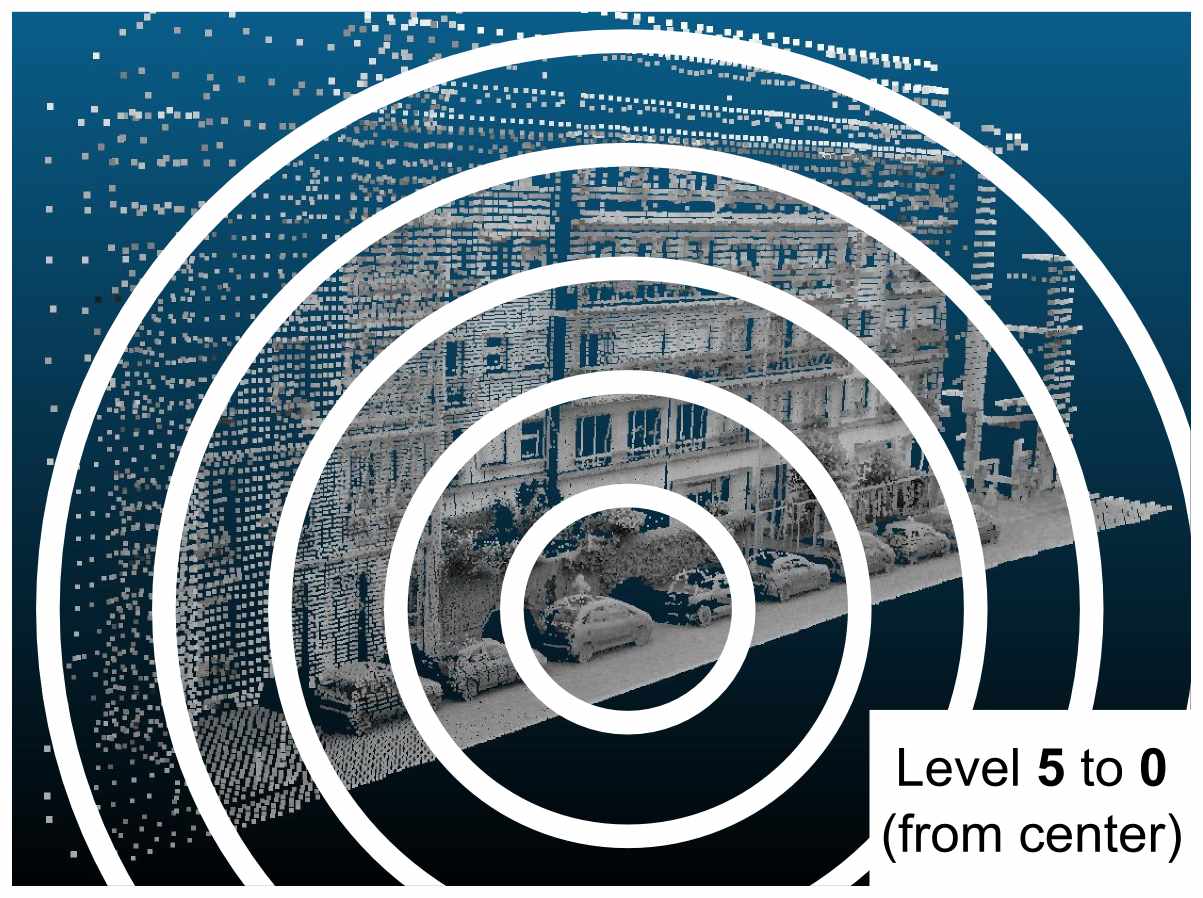}{Schematic example of LOD depending on distance to camera}{lod.fig:lod-dist-to-camera}
		 \\
		 As seen before (See Section \ref{lod.point-cloud-server-troughput}), the point cloud server is fast enough for an internet connection, but is currently slower than a file-based points streaming. Thus for the moment LOD stream is interesting only when bandwidth is limited.
		 
		 The main limitation of this streaming approach is that even when only one point of a patch is displayed, the PCS has to read the whole point from disk, which slows the point retrieval at coarse LOD, when viewing the whole point cloud for instance.
		 On the opposite browsing is pleasantly fast when close enough to points, so that few patches are read from disk, and many points are used.
	
	\subsection{Excessive Density detection and correction}
		\subsubsection{Fast detection}
		Density abnormality detection at the patch level offer the great advantage of avoiding to read points. This is the key to the speed of this method. We don't know any method that is as fast and simple.
		\\
		The limitations stems from the aggregated nature of patch. the number of points per patch doesn't give the density per point, but a quantized version of this per patch.
		So it is not possible to have a fine per point density.

		\subsubsection{Simple correction}
		The correction of density peak we propose has the advantage of being instantaneous and not induce any data loss.
		It is also easy to use as safeguard for an application that may be sensible to density peak : the application simply defines the highest number of points \per \cubic \meter it can handle, and the Point cloud server will always output less than that.
		\\
		The most important limitation this method doesn't guarantee homogeneous density, only a maximum density.
		For instance if an application requires 1000 points \per \cubic \meter for ground patches, all the patches must have more than 1000 points, and patch must have been ordered with MidOc for level 0 to at least 5 ($4^5=1024$). 
		The homogeneous density may also be compromised when the patch are not only split spatially, but with other logics (in our case, points in patch can not be separated by more than 30 seconds, and all points in a patch must come from the same original acquisition file).



\section{Conclusion} 
	Using the Point Cloud Server, we propose a new paradigm by separating the spatial indexing and LOD scheme. Subdivision of point clouds into groups of points (patches) allows us to implicitly store LOD into the order of points rather than externally. 
	After an ordering step, exploiting this LOD does not require any further computation. 
	We propose an geometrical ordering (MidOc) based on the closest point to octree cell centre that produces reliable LOD, successfully used for visualization or as a service for other processing methods (density correction/reduction).
	By also performing intra-level dedicated ordering, we create LOD that can be used partially and still provide good coverage.



	\section{Bibliography}  
	\bibliography{./implicit_LOD} 

%
%
%
%

\end{document}